\documentclass[
reprint,
twocolumn,
superscriptaddress,nofootinbib,
preprintnumbers,
nobibnotes,
 amsmath,amssymb, 
 aps, prl,
 longbibliography,
]{revtex4-1}

\usepackage{cancel}
\usepackage{accents}
\usepackage{mciteplus,slashed}
\usepackage{amssymb,cancel,amsmath}
\usepackage{dcolumn}
\usepackage{bm}
\usepackage{soul}
\usepackage[caption=false]{subfig}
\usepackage{appendix}
\usepackage{physics}
\usepackage{booktabs}
\usepackage{comment}
\usepackage[T1]{fontenc}	
\usepackage{csvsimple}
\usepackage[normalem]{ulem}
\usepackage[bookmarksnumbered=true]{hyperref} 
\hypersetup{
    colorlinks = true,
    linkcolor = blueCERN,
    anchorcolor = blueCERN,
    citecolor = blueCERN,
    filecolor = blueCERN,
    urlcolor = blueCERN
    }
\usepackage[section]{placeins}
\usepackage[capitalise]{cleveref}
\usepackage{booktabs}
\usepackage{graphicx}
\usepackage{mathrsfs}
\graphicspath{{Figures/}}
\AtBeginDocument{
\heavyrulewidth=.08em
\lightrulewidth=.05em
\cmidrulewidth=.03em
\belowrulesep=.65ex
\belowbottomsep=0pt
\aboverulesep=.4ex
\abovetopsep=0pt
\cmidrulesep=\doublerulesep
\cmidrulekern=.5em
\defaultaddspace=.5em
}
\usepackage[dvipsnames]{xcolor}
\usepackage[normalem]{ulem}
\usepackage{fontawesome} 
\usepackage[force]{feynmp-auto}
\usepackage{bbm}
\def\iu{\mathrm{i}}
\def\e{\mathrm{e}}
\def\Ni{N}

\definecolor{amber}{rgb}{1.0, 0.75, 0.0}
\definecolor{darkturquoise}{rgb}{0.0, 0.81, 0.82}
\definecolor{mediumaquamarine}{rgb}{0.4, 0.8, 0.67}
\definecolor{coralred}{rgb}{1.0, 0.25, 0.25}
\definecolor{blueCERN}{HTML}{0033A0}

\newcommand{\uchi}{u_\chi}

\definecolor{interorange}{RGB}{1.0,0.3098,0}
\definecolor{forestgreen}{HTML}{228B22}

\begin{document}

\title{Fluctuations in atom interferometers as a new tool for dark matter}

\author{Clara Murgui}
\email{clara.murgui@cern.ch}
\affiliation{Theoretical Physics Department, CERN, 1 Esplanade des Particules, CH-1211 Geneva 23, Switzerland}
\author{Ryan Plestid}
\email{ryan.plestid@cern.ch}
\affiliation{Theoretical Physics Department, CERN, 1 Esplanade des Particules, CH-1211 Geneva 23, Switzerland}

\date{\today}

\preprint{CERN-TH-2026-024}

\begin{abstract}
 We propose the use of the super-binomial variance in the count rate of an atom interferometer as a novel signature of dark matter. We show that the dark matter induced shift in this observable is enhanced by $N$, the number of atoms used per run of the interferometer, and therefore offers sensitivity that is enhanced by orders of magnitude relative to an independent-atom estimate. As an application, we consider dark matter that interacts with electrons, protons, and/or neutrons, via a long-range Yukawa interaction and new constraints on strongly interacting dark matter that thermalizes in the overburden of conventional direct detection experiments. We find that searches for super-binomial variance extend, and complement, existing atom interferometer observables;  they are well suited to search for both short- and long-ranged forces. 
\end{abstract}

 \maketitle

\vspace{-16pt}

\section{Introduction} 
Dark matter direct detection requires new ideas. An underexplored region of parameter space involves dark matter interacting with matter via small momentum transfers that lie below the $\sim {\rm keV}$ thresholds of noble-element direct detection experiments, and even below the $\sim 0.1 \, {\rm eV}$ thresholds of recently developed semiconductor devices~\cite{Cirelli:2024ssz,Kahn:2021ttr,Essig:2024wtj,Zurek:2024qfm}. Models of this type include very light particle-like dark matter~\cite{Hall:2009bx,Kaplan:2009ag,Holdom:1985ag,Boehm:2003hm,Hochberg:2014dra,Essig:2013lka}, ``dark blobs''~\cite{Zhitnitsky:2002qa,Wise:2014jva,Grabowska:2018lnd}, interactions mediated by a feebly coupled light mediator~\cite{Fayet:1980rr,Arkani-Hamed:2008hhe,Feng:2008dz}, or strongly interacting dark matter thermalizing into a slow, warm gas~\cite{Starkman:1990nj}. Atom interferometers offer compelling sensitivity to models of dark matter that are best searched for via very low momentum transfer scattering with ordinary matter~\cite{Riedel:2012ur,Riedel:2016acj,Du:2022ceh,Badurina:2024nge,Murgui:2025unt}.

When viewed through the lens of a particle detector, atom interferometers have several unique properties. They do not count hard scatters; they instead infer small perturbations to the atoms' density matrix via interferometric measurements~\cite{Geiger:2020aeq}.
A consequence of this scheme is that the detectors are {\it effectively} threshold-less;\footnote{Contributions from low momentum transfers, $q$, much less than the inverse of the interferometer arm-separation, $\Delta x$, are however suppressed by powers of $q\Delta x$~\cite{Riedel:2012ur}.} the phase shift and contrast of the measured fringe are sensitive to the entire spectrum of the differential rate.

More precisely, in the single atom limit~\cite{Joos:1984uk,Hornberger_2003}, for an interferometer counting the number of atoms in two ``ports''  labeled by $\pm$, the fraction of counts, $n_+=N_+/(N_+ + N_-)$, in the $+$ port is expected to follow\footnote{In what follows we assume contrast loss is small and expand $\exp[-\epsilon] = 1-\epsilon+ {\cal O}(\epsilon^2)$ for any quantity $\epsilon\ll 1$.} 
\begin{equation}\label{eq:nplus}
    n_+(\phi) = \frac12 \qty[ 1+ (1-s_0) \cos(\phi+\gamma_0) ]~.
\end{equation}
Here, $\phi$ is the phase in the absence of interactions with the environment, $(1-s_0)$ parametrizes the contrast loss, and $\gamma_0$ is an environmentally induced phase shift. These are the two observables accessible through the mean of counts, $n_+$, for a given tuned phase $\phi$ (an experimental input).  Both $s_0$ and $\gamma_0$ are related to the non-trivial evolution of the off-diagonal $\ketbra{L}{R}$ and $\ketbra{R}{L}$ density matrix elements (where $L$ and $R$ label interferometer paths) and encode background effects (i.e., for a pure state, $s_0, \gamma_0 = 0$). Since we are in the single atom limit, coherent enhancements and cloud-structure dependence are neglected, and the contrast loss and phase shift are controlled respectively by~\cite{Joos:1984uk,Hornberger_2003},  
\begin{align}
    s_0 &= \Phi \int \dd t  \int \dd q^2  \frac{\dd \sigma}{\dd q^2} \langle1-\cos [\vb{q}\cdot \vb*{\Delta x}]\rangle_\phi~,\\
    \gamma_0 &= \Phi \int \dd t  \int \dd q^2  \frac{\dd \sigma}{\dd q^2}\langle \sin [\vb{q}\cdot \vb*{\Delta x}] \rangle_\phi~,
\end{align}
where $\langle \ldots \rangle_\phi = \int_0^{2\pi} \dd \phi/(2\pi) \qty[ \ldots ]$, and we have assumed a unidirectional flux of particles $\Phi$, with a differential cross section on atoms $\dd \sigma /\dd q^2$. In general one must average over incident velocities, which can suppress the phase shift. For example, in an isotropic background the induced phase shift $\gamma_0$ vanishes identically after averaging over incident directions. It is also worth noting that these are essentially the totally inclusive cross sections weighted by dimensionless kernels. 

So far, we have discussed the effects of background interactions on the mean count $n_+$ for a single atom. However, realistic experiments involve a cloud of $N \gg  1$ atoms per run. Recent work has shown how coherent enhancements enter these observables, and how these effects are modified by cloud substructure~\cite{Badurina:2024nge,Murgui:2025unt}. Notably, the induced phase shift receives an $N$-enhancement for coherent interactions that do not resolve the cloud substructure, i.e. $q \ll 1/r_c$, being $r_c$ the radius of the cloud.

When each atom can be treated independently, the outcome of the $i^{\rm th}$ atom is a Bernoulli trial with probability $p_i$; this forms the null hypothesis. Therefore, in the absence of environmental interactions, the statistics of a generic atom interferometer follow a generalized binomial distribution~\cite{PhysRevA.47.3554}.\footnote{This holds when the atomic ensemble is prepared in an uncorrelated product state, which is the easiest atomic state to prepare experimentally.} This leads to definite predictions that relate the mean, variance, and other higher moments of the distribution, even in the presence of experimental noise (e.g., laser fluctuations, etc.). 

In this work, we propose a novel method that can be used to search for particle-like dark matter and other elusive particles. The signal consists of super-binomial noise which signals correlations among the atoms. 

Remarkably, this observable is protected from all sources of noise that alter the individual Bernoulli parameters, $\{ p_i\}$, in an uncorrelated manner. Phrased differently, any noise $X$ that leads to fluctuations in the Bernoulli parameter via a function $P(X)$ cannot fake the proposed dark matter signal. 
This property is guaranteed, under the null hypothesis, by the binomial sum variance inequality. 

Consider $N_T$ Bernoulli trials, $\{x_j\}$, with Bernoulli-parameters\footnote{Note that these are not data points but rather predetermined probabilities of success. In an atom interferometer each Bernoulli parameter $p_i$ depends on random laser fluctuations, and can differ slightly atom-by-atom.}, $\{p_i\}$. To connect with the natural observable of an atom interferometer, we will consider the normalized binomial distribution (using the fraction of counts, $x_j/N_T$, rather than $x_i$ as a variable). The expected sample mean is $\bar{p}= 1/N_T \sum_i p_i$. The expected sample variance, $\sigma^2= (1/N_T^2) \sum_i p_i (1-p_i)$, is then given by
\begin{equation}\label{eq:var0}
    \sigma^2 = \frac{1}{N_T} \bar{p}(1-\bar{p}) - \frac{1}{N_T}\mathfrak{s}^2 \leq \frac{1}{N_T} \bar{p}(1-\bar{p}) ~,
\end{equation}
where $\mathfrak{s}^2 \equiv (1/N_T) \sum_i (p_i-\bar{p})^2 \geq 0$ is computed from the scatter of the input Bernoulli parameters. In essence, if using $\bar{p}$ to predict the variance, one will always systematically {\it overestimate} the true variance. Therefore, any observed variance that is larger (in a statistically significant sense) than $\bar{p}(1-\bar{p})/N_T$ is a signal of correlations among the atoms. As we will now describe, dark matter can induce such correlations and therefore a search for super-binomial noise may be interpreted as a search for dark matter.

\section{Dark matter induced noise} 
The source of super-binomial noise in the presence of interactions between an atom interferometer and the environment stems from the fact that atom interferometers do not perform measurements one atom at a time. Instead, $N_i$ atoms are used in each run of the interferometer, and the experiment is repeated $\aleph$ times. This gives $N_T=\sum_{i=1,\cdots,\aleph} N_i$ total atoms used in the experiment. Although $N_i$ fluctuates due to loss of atoms, the number of $N_+$ and $N_-$ counts can typically be obtained such that $N_i$ is measured {\it a posteriori} and can be taken as fixed. For simplicity we will take $N_i=\Ni$ in what follows, such that $N_T=N\times \aleph$.  

In the presence of $N\geq 2$ atoms, the theory of collisionally induced decoherence changes in a qualitative way~\cite{Badurina:2024nge}. Remarkably, certain subspaces are decoherence free~\cite{Lidar:2013bwk}. For example, labelling the arms of the interferometer by $L$ and $R$ and considering a two-atom system, the off-diagonal entries in the  density matrix $\ket{LR}\bra{RL}$ do not decohere.
This has important consequences for the counting statistics at the time of measurement. 

Consider the $4\times4$ two-body reduced density matrix, $\rho_2$, for a system where $\Ni$ atoms are placed in the superposition $\frac1{\sqrt2}\qty( \ket{L} + \e^{\iu \phi} \ket{R})$. We can build the operator that counts the total number of counts in the $+$ port, $\mathcal{O}_+ = \sum_{i=1, \cdots, \Ni} \ket{+}_i \bra{+}_i~$; the natural observable, defined per shot, is $n_+\equiv \langle \mathcal{O}_+ \rangle / N$  given in \cref{eq:nplus}.  We may then compute the variance of this operator by computing $\langle \mathcal{O}_+^2 \rangle - \langle \mathcal{O}_+\rangle^2$. The expectation values can be written in terms of the corner matrix elements of $\rho_1$ and $\rho_2$ as,  
\begin{align}
    \label{moment-1}
    \langle \mathcal{O}_+\rangle &= \frac{\Ni}{2} \qty( 1+ 2{\rm Re} \mel{L}{\rho_1}{R})~,\\
    \begin{split}
    \label{moment-2}
    \langle \mathcal{O}_+^2\rangle &= \langle \mathcal{O}_+ \rangle + \frac{\Ni(\Ni-1)}{4}\times  \\
     &\hspace{0.025\linewidth}\times \bigg[\frac{3}{2}+
     4{\rm Re} \mel{L}{\rho_1}{R}  
     + 2{\rm Re}\mel{LL}{\rho_2}{RR}\bigg]~,
     \end{split}
\end{align}
where $\rho_1$ and $\rho_2$ are the one and two-body reduced density matrices, respectively. The relevant formulae for the matrix elements in the presence of weak decoherence are
\begin{align}
    {\rm Re} \mel{L}{\rho_1}{R}   &= \frac{(1-\epsilon_1)}2\cos\Theta ~, \\
    {\rm Re} \mel{LL}{\rho_2}{RR} &= \frac{(1-\epsilon_2)}4\cos2\Theta ~,
\end{align}
where we have defined $\Theta \equiv \phi+\gamma_0$, with $\gamma_0$ the dark-matter induced phase shift, and 
\begin{align}
    \epsilon_1 &\equiv \frac12 (\Ni-1) \tau^2 + (s+s_0) ~,\\
    \epsilon_2 & \equiv \frac12(\Ni-2) (2\tau)^2 + 4s+2s_0~.
\end{align}
The parameter $\tau$ is related to dephasing and it is a genuine collective effect~\cite{Badurina:2024nge}. However, since it enters quadratically, we will simply neglect it in what follows since $ \Ni \tau^2 \ll s,s_0$. Nevertheless, it is formally of interest since contrast lost from dephasing scales faster with $\Ni$ than genuine decoherence.

The quantities $s$ and $s_0$ characterize the loss of contrast from coherent and incoherent scattering between dark matter and the atom cloud, respectively. They are structure-dependent, and defined in terms of a time-dependent cloud form factor, $G(\vb{q},t)$, by~\cite{Murgui:2025unt} 
\begin{eqnarray}
    \!\!\!\!\!s  \! \!  &=& \!\!  \Phi \! \int\! \dd t \! \int \! \dd q^2  \frac{\dd \sigma}{\dd q^2} G(\vb{q},t)\langle1-\cos [\vb{q}\cdot \vb*{\Delta x}]\rangle_\phi \ ~,\label{eq:s}\\
    \!\!\!\!\!s_0 \!\!  &=& \!\!  \Phi \! \int \!\dd t \! \int \! \dd q^2  \frac{\dd \sigma}{\dd q^2} \qty[1-G(\vb{q},t)] \langle1-\cos [\vb{q}\cdot \vb*{\Delta x}]\rangle_\phi.
    ~~\label{eq:s0}
\end{eqnarray}

Past proposals have suggested measuring the loss of contrast, as inferred from $n_+(\phi)$, as a probe of dark matter~\cite{Riedel:2012ur,Riedel:2015pxa,Rosi:2017ieh,Du:2022ceh}. This search strategy faces two challenges in atom interferometers: {\it i)} The signal receives no enhancements from the number of atoms in the cloud~\cite{Badurina:2024nge}, and one benefits from large samples only in the reduction of shot noise. {\it ii)} Benign sources of contrast loss (including laser noise, imperfect overlap of interferometer paths, and a host of other sources of noise) present potentially difficult-to-control backgrounds. These all fall in the category of ``$P(X)$'' noise, which induces contrast loss but does not fake the signal of enhanced fluctuations proposed herein.

We instead propose to use a data-driven determination of $\bar{p}$ and to then {\it predict} the variance that is expected under the null hypothesis. In realistic atom interferometers (with a stable laser) the phase noise can be kept small i.e., $\mathfrak{s}^2 \ll \bar{p}(1-\bar{p})$ in \cref{eq:var0}. When computing the increase in the variance due to dark matter interactions with the atom interferometer, one can then approximate $p_i\approx \bar{p}$ as constant. This leads to a prediction for the expected sample-variance, $\sigma^2_{\rm pred}$, of the measurements, $\{ n_+^{(i)}\}$, in the presence of dark matter given by 
\begin{equation}
\begin{split}
    \sigma^2_{\rm pred} (s,s_0)&\equiv \frac{1}{N^2} \left( \langle {\cal O}_+^2 \rangle - \langle {\cal O}_+\rangle^2 \right)\\
    &=
    \frac{\bar{p}(1-\bar{p})}{N}\qty( 1+ \frac{N s_0}{4 \bar{p} (1-\bar{p})}  +2 N s)~, 
\end{split}
\end{equation}
where we have assumed $\mathfrak{s}^2 \ll \bar{p} (1-\bar{p})$ and $N$ constant. This is the variance of the distribution from which counts in the $\pm$ ports are drawn in a $N$-atom run of an interferometer.

Next, consider the small $\bar{p}$ limit. One immediately notices that $\sigma_{\rm pred}^2$ receives corrections proportional to $N s_0 /\bar{p}$ and  $N s$. Therefore, even for couplings that yield a negligible amount of contrast loss, there can be a substantial increase in the sample variance. The signal is maximized by minimizing $\bar{p}$ and maximizing $N$. 

The origin of these enhancements can be understood as follows. In the absence of interactions with the environment, the atoms are in a product state and the expected sample variance is $\bar{p}(1-\bar{p})/N$. Examining \cref{moment-1,moment-2} it becomes clear that a cancellation must take place to ``kill'' the $N^2$ enhanced terms. Indeed this cancellation takes place because $\mel{L}{\rho_1}{R}$ and $\mel{LL}{\rho_2}{RR}$ are closely related to one another precisely because $\rho$ is assumed to be a product state. Interactions between the atoms and dark matter spoil this simple product state picture, and the atoms need not follow binomial statistics. This occurs mechanically in \cref{moment-1,moment-2} because $\epsilon_1$ and $\epsilon_2$ shift the values of $\mel{L}{\rho_1}{R}$ and $\mel{LL}{\rho_2}{RR}$ in a manner that spoils the cancellation discussed above. One is then left with a variance in the counts which scales as $N^2$ (and therefore the variance, $\sigma^2_{\rm pred}$, that is independent of $N$) and is unsuppressed even when $\bar{p}=0$. 

In order to claim a statistically significant signal, at say 95\% confidence level (CL), one requires that the observed sample variance would have a probability of $5\%$ or smaller of arising given an $N$-atom variance of $\sigma^2_{\rm pred}$. This may be estimated setting the background effects to zero, $s,s_0\rightarrow 0$, and explicitly computing {\it the variance of the variance} (VoV) of the normalized binomial distribution. The VoV of the normalized binomial distribution is
\begin{equation}
    \label{vov-null}
    {\rm var}(\sigma^2_0) = \frac{2 \bar{p}^2(1-\bar{p})^2}{N^2} + \frac{\bar{p}(1 - \bar{p})  (1 - 6 \bar{p}(1- \bar{p}) )}{N^3}~.
\end{equation}
Below we describe in detail a protocol to convert {\it anomalous} (dark-matter induced) fluctuations in an atom interferometer into a constraint on dark-matter atom scattering.

 \section{Experimental protocol} 
 Let us assume that an atom interferometer uses $N$ atoms per shot, and takes $\aleph$ shots. During data taking, the mean of each shot, $n_+^{(i)}$, is recorded (e.g., via fluorescence~\cite{RevModPhys.81.1051,Rocco_2014}). The interferometer may be adjusted shot-to-shot to attempt to minimize $\bar{p}$; we discuss some details of this adjustment scheme below. After $\aleph$ shots, a data set $\{n_+^{(i)}\}$ is assembled. The sample mean is computed and used as an estimator for $\bar{p}$
\begin{equation}
    \bar{p}_{\rm obs} = \frac{1}{\aleph} \sum_{i=1}^\aleph n_+^{(i)} ~.
\end{equation}
For simplicity, we will assume that $\bar{p} N_T \geq 100$, such that Gaussian statistics apply. The estimate $\bar{p}_{\rm obs}$ allows for an inference of $\bar{p}$ with a shot-noise limited error $\delta \bar{p} \sim 1/\sqrt{N_T}$, which is negligible.\footnote{As discussed below, we assume a laser phase noise of $\delta \phi \sim {\cal O}(10)$ mrad, and $\aleph \gg 1$ shots. Under these conditions, laser phase noise constitutes the dominant source of uncertainty.}

Next, the observed sample-variance,
\begin{equation}
    \sigma^2_{\rm obs} = \frac{1}{\aleph}\sum_{i=1}^\aleph (n_+^{(i)}-\bar{p}_{\rm obs})^2~,
\end{equation}
is compared to the prediction\footnote{Here we treat the variance of counts for an $N$-atom run as the observable. Each run of the interferometer is identically distributed and independent, whereas the atoms in a single run are not (because of dark-matter induced correlations).} $\sigma^2_{\rm pred} = \bar{p}(1-\bar{p})/N$ under the null hypothesis, where $\bar p \approx \bar p_{\rm obs}$. If the difference exceeds three standard deviations, then {\it evidence} is obtained for dark matter. Alternatively, if the difference is small, then we may interpret this null-result as excluding couplings $\alpha_{\rm DM}$ larger than some threshold. Demanding $\aleph \gg 1$, and appealing to the central limit theorem, we may use Gaussian statistics and obtain a 95\%-CL by solving,
\begin{equation}
    \sigma^2_{\rm pred}(s,s_0) -\sigma^2_{\rm obs}   \leq   2 \sqrt{{\rm var}(\sigma^2_0)/\aleph}~,
\end{equation}
with the VoV, ${\rm var}(\sigma^2_0)$, computed under the null-hypothesis as given in \cref{vov-null}. 

We compute a 95\%-CL, assuming the experiment sees precisely what is expected under the null hypothesis, a sample variance of $\sigma_{\rm obs}^2=\bar{p}(1-\bar{p})/N$, and that $\mathfrak{s}^2$ is negligible\footnote{For a randomly fluctuation phase $\Theta$ close to $0$, one can show that $\bar{p} \simeq \frac14\langle \Theta^2\rangle$ and $\mathfrak{s}^2 \simeq \frac1{16}(\langle \Theta^4 \rangle - \langle \Theta^2 \rangle^2)$, such that $\mathfrak{s}^2 \ll \bar{p}(1-\bar{p})$.} relative to $\sqrt{{\rm var}(\sigma^2_0)/\aleph}$. This 95\%-CL then serves as our working definition of the {\it sensitivity} of an atom interferometer. 

Let us discuss the scaling of the sensitivity with $N$ and $\aleph$. Clearly, as $\aleph$ increases, the sensitivity improves like $\sqrt{\aleph}$. As $N$ increases, the standard deviation of the variance, $\sqrt{{\rm var}(\sigma^2_0)}$, decreases like $1/N$. The pure-binomial  (i.e., the null hypothesis) variance also drops;  however, the total predicted variance in the presence of dark matter does not. Instead, $\sigma^2_{\rm pred} \rightarrow \frac14 s_0 + 2\bar{p}(1-\bar{p})s$ due to the $N$-enhanced fluctuations induced by dark matter atom scattering. The sensitivity can be {\it further} enhanced by tuning the interferometer towards $\bar{p}\rightarrow 0$, which suppresses the VoV but not $\sigma^2_{\rm pred}$. Putting all of these factors together leads to a sensitivity to the incoherent scattering regime that scales as 
\begin{equation}\label{scaling}
    s_0 \leq \frac{8 \sqrt{2}}{N}  \frac{1}{\sqrt{\aleph}}\bar{p} \left( 1 + \frac{1}{2\bar{p} N}\right)^{1/2}~. 
\end{equation}

Let us now estimate what a realistic value for $\bar{p}$ could be. Imagine an interferometer which after a single shot returns a sample-average of $\bar{p}_1 > 0$. The experimentalist may then tune their laser to try to reduce the sample average on the next shot $\bar{p}_2$. Clearly, there is a fundamental limit of $\bar{p}$ due to noise. In particular, any phase-instability from shot-to-shot cannot be mitigated by a large number of samples because $\bar{p}>0$ is positive definite, and so will never {\it average down}. Moreover, unlike in the case of a phase measurement, the noise cannot be reduced by measuring relative phase difference between two interferometers. Taking phase noise in the $1-30~{\rm mrad}$ range, we expect that $\bar{p} \sim 10^{-6}-10^{-4}$ is possible since $\bar{p} \sim \frac14 \langle \Theta^2 \rangle$ scales quadratically with the phase for $\Theta\ll 1$. We expect this to be limited by instrumentation noise (vibrations, laser fluctuations, etc.) rather than by shot-noise. In the following,  we label the instrumentation noise on the phase as $\delta \phi$.

This constraint on super-binomial fluctuations may be translated into a projected sensitivity to the strength of the scattering between dark matter and the atoms. To place \cref{scaling} into context, recall that current atom interferometers use $N \sim 10^6$ atoms. Future proposals suggest (optimistically) that $N\sim 10^{10}$ atoms and a phase noise power spectral density of $10^{-5}~{\rm rad}/\sqrt{\rm Hz}$ could be achieved \cite{AEDGE:2019nxb}. This power-spectral density, however, assumes a reference interferometer to suppress common sources of noise. For example laser noise is mitigated by mode-locking the laser between the two interferometers. For the present search, these sources of noise suppression are ineffective since we are only interested in $\bar{p}_{\rm est}$. We therefore expect that the phase noise can be reduced to $1-30~{\rm mrad}$, but not further. This leads to the estimate 
\begin{equation}
    \label{eq:s0-bound}
    s_0 + 8\bar{p} s \leq  3 \times 10^{-17} \bigg(\frac{10^{10}}{N}\bigg) \bigg(\frac{10^5}{\aleph}\bigg)^{1/2}\bigg( \frac{\bar{p}}{10^{-5}}\bigg)~.
\end{equation}
Using realistic values (phase noise of $\delta \phi ={\cal O}(1-10)$ mrad, and a large atom cloud $N \gtrsim 10^6$) gives   $\bar p N  \gg 1$ and the second term in \cref{scaling} can be neglected.
The term $\bar{p} s$ is typically negligible relative to $s_0$ except in the case of very light mediators which enhance the coherent regime relative to the incoherent one.

Before proceeding to phenomenology we briefly comment on potential backgrounds. As we have emphasized above, $P(X)$ noise cannot fake the signal we have proposed. However, correlations between the atoms (specifically correlations within the internal-state Hilbert space) can serve as a bona fide background. Much like backgrounds in direct detection experiments, correlations between atoms can be mitigated with good experimental design and {\it in~situ} measurements.

In the following, we illustrate the potential of the proposed atom interferometer observable with two representative dark-matter scenarios.

\section{Yukawa-mediator dark matter} 
Let us consider, first, a concrete model often adopted in the context of direct detection with atom interferometers \cite{Riedel:2016acj,Du:2022ceh,Murgui:2025unt}. We assume that dark matter interacts with atoms via a long range Yukawa potential $V(r)= A \, \alpha_{\rm DM} \, \e^{-m_\phi r}/r$ where $A=N_p+N_n$ counts the number of nucleons in the nucleus.  
We use natural units ($c=\hbar = 1)$. The differential dark matter-nucleus cross section is given by 
\begin{equation}\label{eq:DMxsec}
    \dv{\sigma}{q^2} = 4\pi A^2  \, \alpha_{\rm DM}^2 \frac{1}{\uchi^2} \qty(\frac{1}{q^2+m_\phi^2})^2~,
\end{equation}
where $q=|\vb{q}|$ is the momentum transfer, and $\uchi$ is the dark matter velocity. 
Microscopically this interaction arises from the exchange of a light boson of mass $m_\phi$ that couples to neutrons and protons with equal charge.

\begin{figure}[h]
    \centering
 \includegraphics[width=\linewidth]{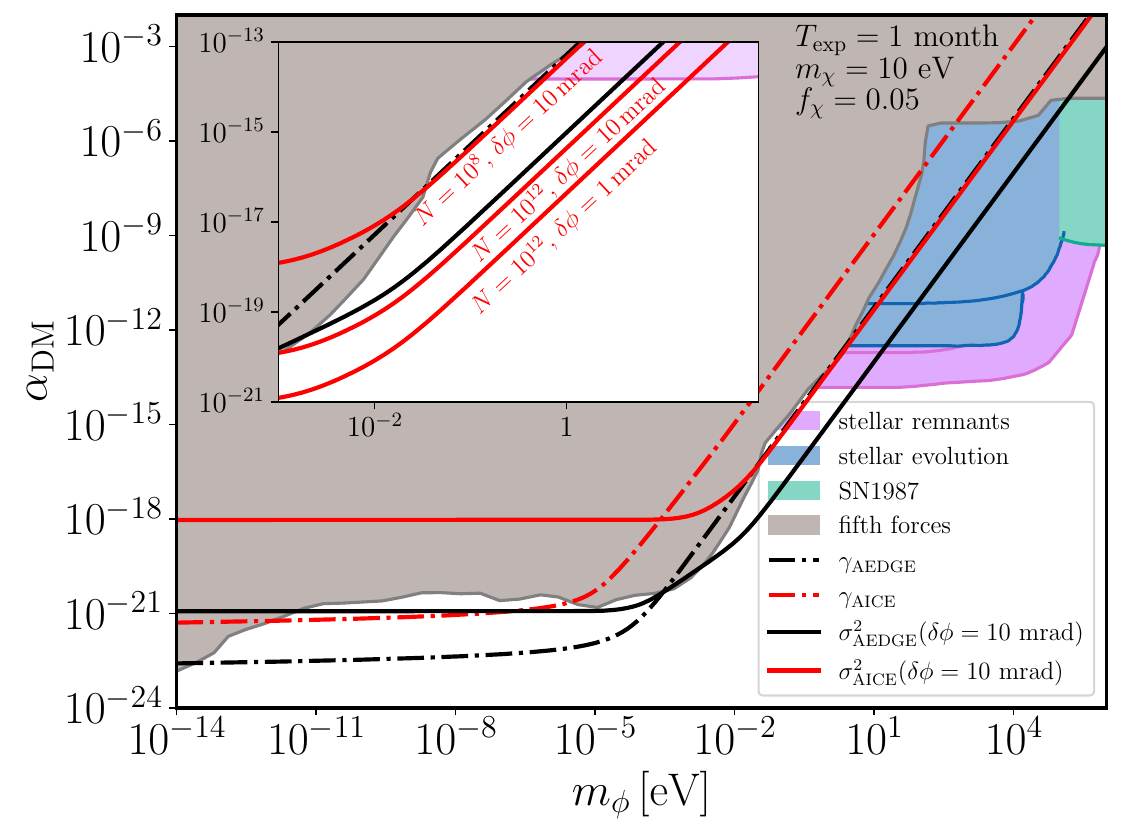}
    \caption{Sensitivity projections for anomalous phase shift (dot-dashed curves) and super-binomial fluctuations (solid curves) searches with atom interferometers. For all curves, we take $T_{\rm exp} = 1~{\rm month}$, the dark matter coupling to the mediator to be $y_\chi = \sqrt{4\pi}$, an atomic species with $\text{A}= 87$ nucleons (e.g.,$^{87}\text{Rb}$ or $^{87}\text{Sr}$), and a subcomponent dark matter with $\rho_\chi = 0.05 \, \rho_{\rm DM}$ (in order to evade the strong self-interaction dark matter constraints at low dark matter masses). The black and red curves adopt the atom interferometer benchmarks quoted for the proposals AEDGE~\cite{AEDGE:2019nxb} and AICE~\cite{Baynham:2025pzm}, respectively. For AEDGE, a proposed space-based mission, a cloud of $N=10^{10}$ atoms is assumed, with $\mathcal{T}_{1/2}=600\text{ s}$, $\Delta x = 0.9 \text{ m}$, and $r_c = 4~{\rm mm}$. AICE, the proposal based at CERN, assumes $N=10^8$ atoms and $\mathcal{T}_{1/2}=1.5~{\rm s}$, $\Delta x = 50~{\rm m}$, and $r_c = 100~\mu{\rm m}$. The solid lines display the reach of the new observable proposed in this paper: super-binomial fluctuations of the atom interferometer. For comparison, we also indicate with dot-dashed curves the sensitivity of the coherently enhanced anomalous phase-shift following Ref.~\cite{Murgui:2025unt}. The inner plot zooms on a patch of parameter space and shows the sensitivity from super-binomial fluctuations adopting AICE as benchmark with different number of atoms and instrumental phase noises, as indicated in the figure. The gray shaded area shows the parameter space ruled out by fifth force searches~\cite{Murata:2014nra}. In several cold colors we show the excluded parameter space from the stellar astrophysical constraints on light mediators: bounds from SN1987~\cite{Hardy:2024gwy} in green, from stellar evolution stages (red giant and horizontal branch stars~\cite{Hardy:2016kme}) in blue, and from stellar remnants (neutron stars~\cite{Fiorillo:2025zzx} and white dwarfs~\cite{Bottaro:2023gep}) in purple.}
\label{fig:LDM}
\end{figure}

\Cref{fig:LDM} shows, in solid lines, the constraints derived from the non-observation of super-binomial noise within a control in the uncertainty of the input phase of $\delta \phi = 10~ {\rm mrad}$  on a $5\%$ subcomponent dark matter ($f_\chi = 0.05$) with $m_\chi = 10$ eV. The solid black and red lines in the main plot adopt the benchmarks quoted in the atom interferometer proposals AEDGE~\cite{AEDGE:2019nxb} (space mission) and AICE~\cite{Baynham:2025pzm} (LHC shaft at CERN), respectively.

For the incoherent decoherence, $s_0$, there are three relevant scales, $p_\chi$, $m_\phi$ and $\Delta x$. When $m_\phi \gg p_\chi$, \cref{eq:DMxsec} is a contact interaction and $s_0$ is dominated by the high-$q^2$ bins. By applying the Riemann-Lebesgue lemma,\!\footnote{For particle-like dark matter, $(\Delta x)^{-1} \ll p_\chi$.} the quantity $s_0$ can be interpreted as the number of scattering events, $s_0 \simeq \Phi \, \mathcal{T}  \sigma$.
In \cref{fig:LDM}, the benchmark $m_\chi = 10 ~{\rm eV}$ is adopted, and one can read the $m_\phi^{-4}$ dependence of $s_0$ for mediator masses heavier than $p_\chi \sim 10~{\rm meV}$.
For  $m_\phi \ll p_\chi$, the distance between the clouds also matters. In particular, for $m_\phi \gg (\Delta x)^{-1}$, $s_0$ can still be interpreted as the total number of scatters and the scaling with the mediator mass goes like $m_\phi^{-2}$, while for $m_\phi \ll (\Delta x)^{-1}$, the differential cross-section gets penalized by the lower-$q$ cutoff implemented by the decoherence kernel so that $s_0$ has a flat dependence with $m_\phi$ (see \cref{fig:LDM}).

We incorporate the dark matter velocity distribution by taking averages with respect to a dark matter momentum distribution. We simply replace $\uchi^{-1}\rightarrow \langle \uchi^{-1}\rangle$. This quantity will modulate over a sidereal day and can be used to distinguish a dark matter induced signal from other benign sources of increased sample variance.

Benchmarking with realistic atom interferometer parameters, we find a sensitivity to the dark-matter nucleon coupling of 
\begin{equation}
\begin{split}
    \label{eq:alphaDM}
   & \alpha_{\rm DM}< 1.1\times 10^{-14} \left(\frac{10^8}{N}\right)^{\tfrac{1}{2}}\left(\frac{\delta \phi}{10 \text{ mrad}}\right) \left(\frac{87}{A}\right)\\
    &  \times \left(\frac{1\text{ s}}{\mathcal{T}_{1/2}}\right)^{\tfrac{1}{4}}\left(\frac{30 \text{ days}}{T_{\rm exp}}\right)^{\tfrac{1}{4}} \left(\frac{m_\phi}{1 \text{ eV}}\right)^{2} \left(\frac{10 \text{ eV}}{m_\chi}\right)^{\tfrac{1}{2}}~.
\end{split}
\end{equation}
The curves corresponding to the new observable proposed in this paper are displayed together with the complementary bounds from the coherently-enhanced anomalous phase shift, $\gamma$, discussed in Refs.~\cite{Badurina:2024nge,Murgui:2025unt}, shown in dashed-dotted lines. In the inner subplot, dashed red curves show the super-binomial fluctuations with different number of atoms per run to illustrate the $1/\sqrt{N}$ scaling discussed above, as well as optimistic instrumentation phase noises $\delta \phi$.

 \section{Strongly interacting dark matter} 
 Let us discuss a second phenomenological application of the above ideas. Namely, searching for dark matter cross sections that are so large that dark mater cannot penetrate conventional direct direction overburdens. Dark matter direct detection often focuses on GeV-scale dark matter with incredibly small, $\sigma_{\chi N} \lesssim 10^{-40}~{\rm cm}^2$(e.g.,~\cite{XENON:2019izt}), dark matter-nucleon cross sections. At sub-GeV masses, the last decade has seen rapid progress in semiconductor detectors targeting dark matter-electron cross sections~\cite{Essig:2011nj,Essig:2015cda,Dolan:2017xbu} down to $\sigma_{\chi e} \lesssim 10^{-36}~{\rm cm}^2$ (e.g.,~\cite{SENSEI:2023zdf}); the Migdal effect~\cite{Ibe:2017yqa} also offers some sensitivity to dark matter nucleon interactions for $m_\chi\lesssim 1~{\rm GeV}$ (e.g.,~\cite{DarkSide:2022dhx}). Other condensed matter and molecular systems allow for ultralow thresholds that can probe dark matter masses down to $m_\chi \gtrsim 1~{\rm keV}$~\cite{Hochberg:2015pha,Hochberg:2016sqx,Knapen:2017ekk,Essig:2024wtj}. 

All of these experiments, however, have a {\it ceiling}. For dark matter-nucleon (or electron) cross sections above some critical value, the dark matter cannot penetrate the experiment's overburden. Thus, counterintuitively, large dark matter-nucleon (and dark matter-electron) cross sections also represent a direct detection challenge. Above $m_\chi \gtrsim 1~{\rm GeV}$, a reinterpretation of the rocket-based X-ray Quantum Calorimeter (XQC)~\cite{Erickcek:2007jv} provides constraints on the dark matter-nucleon cross section, $\sigma_{\chi N}$. When plotting the different constraints on $\sigma_{\chi N}$, we consider a reinterpretation of the XQC data with an optimistic~\cite{Erickcek:2007jv} and a more realistic~\cite{Mahdawi:2018euy} thermalization efficiency of the nuclear recoil energy. Below $m_\chi \lesssim 1~{\rm GeV}$, almost no direct detection constraints exist at large cross sections; the super-binomial noise signal proposed above, however, can naturally fill this gap. 

In the context of strongly-interacting dark matter (SIDM) the observable benefits from the following properties: {\it i)} SIDM often thermalizes in the Earth and/or atmosphere, which causes the induced phase shift, $\gamma_0$ to vanish; the decoherence rate $s_0$ does not suffer from this deficiency.  {\it ii)} Having thermalized, SIDM has $\uchi \sim \sqrt{T_\oplus/ m_\chi}$ with $T_\oplus \sim 30~{\rm meV}$ the temperature of the Earth. Atom interferometers, being thresholdless, can detect this thermalized population, while conventional direct detection experiments cannot. {\it iii)} Furthermore, the {\it traffic jam} mechanism can lead to $n_\chi^{\oplus} \gg n_\chi^{\rm vir.}$ with $n_\chi^{\rm vir.}$ the virial number density. For cross sections that scale like $1/\uchi^4$ (Yukawa-like force) or $1/\uchi^2$ (dipole interactions) this can lead to remarkably large rates of induced decoherence due to the non-relativistic enhancements. 

Since this effect, although enhanced by $N^2$, stems from {\it incoherent scattering}, it can produce competitive constraints even for contact interactions (which do not favour forward scattering). For long-range interactions, such as those mediated by a Yukawa-like force or dipole interaction, one should also consider the coherent contribution; however, the incoherent regime always contributes at least an $O(1)$ fraction of the total effect. Therefore, the primary consequence of long-range interactions are to induce the velocity scalings mentioned above, which can then enhance the sensitivity to feeble dark matter couplings. 

\begin{figure}
    \centering
    \includegraphics[width=\linewidth]{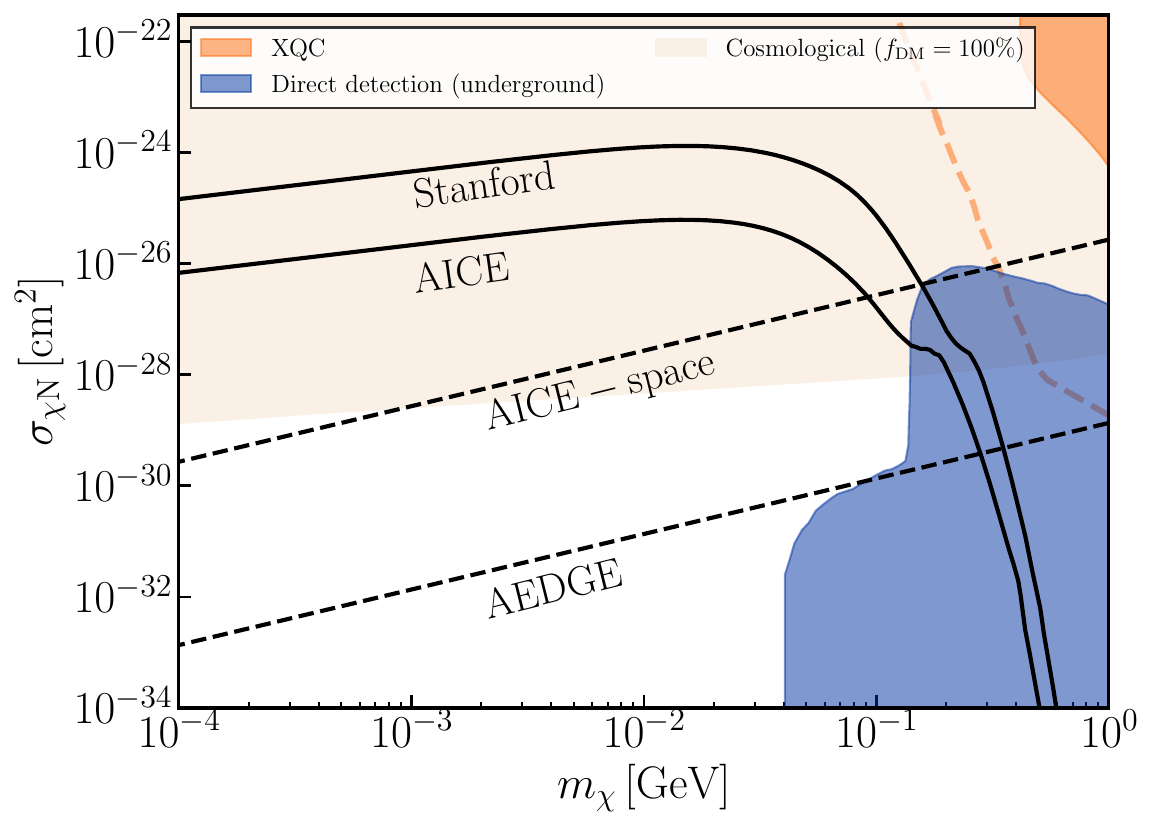}
    \caption{Sensitivity curves for strongly interacting dark matter scattering off nucleons. Solid lines show the projected reach of the current Stanford atom fountain~\cite{Overstreet:2021hea,Asenbaum:2020era}, located at the Earth’s surface, and of the proposed atom interferometer AICE~\cite{Baynham:2025pzm}, planned to operate about 100 m underground (at the PX46 access shaft to the LHC). Dashed lines show the projected sensitivity of prospective space-based atom interferometers, including the proposed AEDGE~\cite{AEDGE:2019nxb} and a space-based (fictitious) analogue of AICE~\cite{Baynham:2025pzm}. These sensitivities probe regions of parameter space not accessible to other direct-detection constraints, including underground nuclear-recoil experiments (shown in blue: DarkSide-50~\cite{DarkSide:2022dhx}, CRESST~\cite{CRESST:2019jnq,CRESST:2019axx,CRESST:2017ues}, XENON-1T~\cite{XENON:2017vdw,PhysRevLett.123.251801}, and CDEX-10~\cite{PhysRevLett.129.221802}), and the rocket-based XQC experiment~\cite{McCammon_2002} (orange). On the reinterpretation of the XQC data, we show with a dashed orange curve the interpreted bounds on $\sigma_{\chi N}$ assuming an optimistic $100\%$ efficiency for thermalization of the nuclear recoil energy in the sub-keV range~\cite{Erickcek:2007jv}, and with a solid orange curve the reinterpreted bounds adopting a more realistic thermalization efficiency~\cite{Mahdawi:2018euy}. For underground direct-detection experiments we adopt the upper bounds ({\it ceilings}) from Refs.~\cite{Emken:2018run,Kavanagh:2017cru}, and estimate the ceiling for DarkSide-50. Areas covered with fainted beige indicate cosmological constraints~\cite{Buen-Abad:2021mvc}
As discussed in the main text, these are model dependent and may be evaded, for example, if dark matter constitutes only a subcomponent or exhibits velocity-dependent scattering.}
    \label{fig:SIDM}
\end{figure}

As an application we consider the least favorable scenario, in which the dark matter-atom\footnote{The dark matter-atom interactions can emerge from either dark matter-nucleon or dark matter-electron couplings.} cross section is velocity independent. For simplicity we will assume that $\sigma_{\chi\, ^{87}{\rm Rb}} = A_{^{87}{\rm Rb}}^2 \sigma_{\chi N}$, with $\sigma_{\chi N}$ the dark matter-nucleon cross section. We consider both an atom interferometer on the surface of the Earth (such as the Stanford atom fountain~\cite{Overstreet:2021hea,Asenbaum:2020era} or AICE~\cite{Baynham:2025pzm}) and in outer space (such as AEDGE~\cite{AEDGE:2019nxb}). The former of these setups probes the thermalized dark matter population near the Earth's surface, while the latter probes the standard virialized dark matter halo. 

In addition to direct detection constraints, there also exist cosmological and astrophysical constraints on dark matter-nucleon interactions. These come with their own systematic uncertainties and caveats; however, they provide important context. Constraints include those that stem from the CMB~\cite{Gluscevic:2017ywp,Buen-Abad:2021mvc}, from suppression of the linear power spectrum and its eventual influence on the Milky Way satellite distribution~\cite{Nadler:2019zrb,Buen-Abad:2021mvc}, and from the Lyman-$\alpha$ forest~\cite{Xu:2018efh,Buen-Abad:2021mvc}. They can be significantly weakened if dark matter constitutes a subcomponent of the total dark-matter density.

We also highlight the constraints from cosmic ray upscattered dark matter (CRUD)~\cite{Bringmann:2018cvk,Cappiello:2018hsu,Krnjaic:2019dzc,Cappiello:2019qsw}. Notably, these constraints stem from {\it relativistic} kinematics and the mapping between the CRUD cross section (used both to generate the up-scattered flux and to detect it on Earth) and that probed in the atom interferometer setup is model dependent. For a contact interaction, the cross section at higher energies is much larger. If the cross section is dominated by the exchange of a massive mediator, whose mass is lower than the typical center of mass energy in cosmic ray dark matter collisions, then the cross section can be {\it smaller} at large energies. Mapping this cross section onto that measured in an atom interferometer can then raise or lower the location of the CRUD constraints when expressed in the $(m_\chi, \sigma_{\chi N})$ plane. For simplicity, we therefore omit them when presenting projected atom interferometer sensitivities to SIDM, although they should be included in any model-specific analysis. 

In \cref{fig:SIDM} we show projected sensitivities for an atom interferometer on the surface of the Earth and in space. We use AICE~\cite{Baynham:2025pzm} (a proposed ground-based experiment at CERN) and {\it AICE in space} (a fictitious space-based experiment with the same parameters as AICE) to illustrate the role of the thermalized dark matter population on Earth. We additionally show projected constraints that could be obtained from the (existing) Stanford atom fountain~\cite{Overstreet:2021hea,Asenbaum:2020era}, and the proposed space-based AEDGE mission~\cite{AEDGE:2019nxb}. We use the one-dimensional (radial) hydrodynamic formalism outlined in Refs.~\cite{Neufeld:2018slx,Ema:2024oce} to compute the thermalized number density on Earth (we use the preliminary Earth reference model~\cite{Dziewonski:1981xy}, the Earth profile temperature from Ref.~\cite{etde_22487687}, and the NRLMSIS-00 model the density, temperature, and composition of the atmosphere~\cite{Picone2002NRLMSISE00}). 

Atom interferometers, even on Earth, can provide the first direct detection constraints on dark matter cross sections larger than $\sim 10^{-30}~{\rm cm}^2$ (depending on the dark matter mass). Since space-based missions such as AEDGE are already planned, constraints can be obtained even at light masses where traffic jam populations on Earth are relatively modest. A further investigation of dark matter direct detection prospects below $m_\chi \sim 1~{\rm MeV}$ is a natural future direction. We emphasize that these constraints are much less model dependent than cosmological and astrophysical constraints. 

\section{Conclusions}
We have proposed a novel strategy to search for dark matter via higher statistical moments of atom interferometer data. The enhanced statistical fluctuations due to dark matter becoming entangled with the cloud of atoms results in super-binomial fluctuations which {\it cannot be faked} by laser noise or any other stochastic variable that modifies the individual atoms' Bernoulli parameter. The only source of backgrounds are genuine many-body correlations among the atoms (in the internal-state Hilbert space), and these can be mitigated and tested-for experimentally. This is analogous to the mitigation of radioactive backgrounds in conventional direct detection experiments, which serve as irreducible background but can be directly measured, quantified, and controlled by the collaboration. 

The physical origin behind the $N$-enhanced scaling is the induced correlations between the atoms from their interactions with the dark matter. These correlations allow for super-binomial fluctuations in the atom counts. The proposed technique exhibits a scaling of $\sim \sqrt{N} \times \sqrt{N_{T}} \times 1/\bar{p}$ for the cross-section sensitivity, where $N$ is the number of atoms in each run of the interferometer, $N_T= N\times \aleph$ the total number of atoms in the lifetime of the experiment, and $\bar{p}$ the average probability of an atom to appear in the $+$ port. This offers a gain in sensitivity of $\sim \sqrt{N}/\bar{p}$ over the naive shot-noise estimate of $\sqrt{N_T}$. Since $\bar{p}$ is numerically close to $1/\sqrt{N}$ for the setups we consider, this is competitive with the $N \times \sqrt{N_T}$ scaling that is obtained from a coherently enhanced phase shift.

The search for enhanced statistical fluctuations has several advantages over previously proposed atom interferometer observables. Unlike contrast loss, the signal cannot be faked by $P(X)$ noise, and so may offer a more robust probe of dark-matter atom scattering while simultaneously providing $N$-enhanced sensitivity. Unlike the coherently enhanced phase shift, the statistical correlations that are induced by scattering are present even for an isotropic flux of dark matter, and are most naturally suited to probe the regime of {\it incoherent scattering}. As a result, the sensitivity has very different parametric scaling as compared to the phase shift observable previously considered. We have outlined how these properties can be used to search for SIDM (with short-range interactions) and Yukawa-like forces of intermediate range.

\textbf{Acknowledgments:} We thank Graham Kribbs and Marios Galanis for helpful feedback and discussions. We thank Markus Ardnt, Stefan Gerlich and Edoardo Vitagliano for useful comments and suggestions, and the University of Vienna for their hospitality during RP's visit.

\bibliography{biblio.bib}

\end{document}